# Leptonic and Semileptonic Decays of Charm and Bottom Hadrons

Jeffrey D. Richman

Department of Physics, University of California, Santa Barbara, CA 93106
richman@charm.physics.ucsb.edu

Patricia R. Burchat

Department of Physics, Stanford University, Stanford, CA 94305
pat@slac.stanford.edu

## Abstract

We review the experimental measurements and theoretical descriptions of leptonic and semileptonic decays of particles containing a single heavy quark, either charm or bottom. Measurements of bottom semileptonic decays are used to determine the magnitudes of two fundamental parameters of the standard model, the Cabibbo-Kobayashi-Maskawa matrix elements $V_{cb}$ and $V_{ub}$. These parameters are connected with the physics of quark flavor and mass, and they have important implications for the breakdown of CP symmetry. To extract precise values of $|V_{cb}|$ and $|V_{ub}|$ from measurements, however, requires a good understanding of the decay dynamics. Measurements of both charm and bottom decay distributions provide information on the interactions governing these processes. The underlying weak transition in each case is relatively simple, but the strong interactions that bind the quarks into hadrons introduce complications. We also discuss new theoretical approaches, especially heavy-quark effective theory and lattice QCD, which are providing insights and predictions now being tested by experiment. An international effort at many laboratories will rapidly advance knowledge of this physics during the next decade. **This file contains only the abstract and the table of contents. The full document is at the URL http://charm.physics.ucsb.edu/papers/slrevtex.ps and contains 168 pages and 47 figures.**

To be published in *Reviews of Modern Physics.*

## CONTENTS